\documentclass[
    ,final            
  ]
  {aipproc}

\layoutstyle{6x9}

\begin{document}

\title{From dense-dilute duality to self duality in high energy evolution}


\author{Alex Kovner}{
  address={Physics Department, University of Connecticut, Storrs, CT 06269, USA}
}

\begin{abstract}
I describe recent work on inclusion of Pomeron loops in the high energy evolution. In particular I show that the complete eikonal high energy evolution kernel must be selfdual.
\end{abstract}

\maketitle


Last year has seen renewed attempts to understand Pomeron loop contributions to the high energy evolution of hadronic cross sections in QCD.
In recent years the study of the high energy scattering has centered around the so called JIMWLK evolution equation \cite{balitsky,JIMWLK,cgc}. It describes the approach of the scattering amplitude to saturation due to multiple scattering corrections on dense hadronic targets, or in the diagrammatic language, the fan diagrams.
The JIMWLK equation however only partially takes into account the processes whereby the gluons emitted in the projectile wave function at an early stage of the evolution, are "bleached" by subsequently emitted gluons, or the so-called Pomeron loops\cite{im1},\cite{ms},\cite{ploops},\cite{kl},\cite{kl2}.

Recently we have calculated corrections to the JIMWLK equation, which take into account some finite density effects in the projectile wave function (or equivalently, resum certain corrections away form the dense limit of the the target)\cite{kl1}.
We have also derived the evolution equation valid for dilute target, which is the opposite limit to that considered in JIMWLK\cite{kl}.
The most striking feature of the two results, is that they appear to be dual to each other. The
improved JIMWLK equation is given by\cite{kl1}
\begin{eqnarray}
\chi^{\rm JIMWLK+}&=&
\frac{1}{2\pi}\int_z\left\{ b^a_i(z,0,[{\delta\over \delta \alpha}])b^a_i(z,0,[{\delta\over \delta \alpha}])+ 
 b^a_i(z,1,[{\delta\over \delta \alpha}])b^a_i(z,1,[{\delta\over \delta \alpha}])
\right.\nonumber\\
&-&2\, \left. b^a_i(z,0,[{\delta\over \delta \alpha}])\left[{\cal P}e^{i\int_{0}^{1} d x^-T^c\alpha^c(x^-,z)}\right]^{ab}b^b_i(z,1,[{\delta\over \delta \alpha}])\right\}
\label{notlarge}
\end{eqnarray}
where $\cal P$ denotes path ordering with respect to $x^-$ and the field $b^a_i$ satisfies the "classical" equation of motion\cite{kl1}.
The low density limit evolution kernel (KLWMIJ)  including the same type of corrections but in the target wave function derived in \cite{kl} is\footnote {Following the work \cite{kl}, the same expression has also been obtained  in  \cite{smith} using the effective action techniques.}: 
\begin{eqnarray}\label{notsmall}
\chi^{\rm KLWMIJ+}=&-&
\frac{1}{2\pi}\int_{z} \{ b^a_i(z,0,[\rho])b^a_i(z,0,[\rho])
+b^a(z,1,[\rho])b^a(z,1,[\rho])
\nonumber \\
&-& 2\,b^a_i(z,0,[\rho])\left[{\cal P}e^{\int_{0}^{1} d x^-T^c{\delta\over\delta\rho^c(x^-,z)}}\right]^{ab}b^b_i(z,1,[\rho])\}\,.
\end{eqnarray}

The two kernels are strikingly similar which suggests an intriguing duality between the high and the low density limits of the evolution kernel.

In this contribution I follow \cite{kl2} and show that indeed the full eikonal kernel for the high energy evolution must satisfy the property of self duality.
The requirement that the evolution of the projectile and the target wave functions has the same functional form coupled with the requirement of Lorentz invariance of the scattering matrix, leads to the condition that the kernel of the evolution $\chi[\rho,{\delta\over\delta\rho}]$ must satisfy
\begin{equation}\label{main}
\chi[\alpha,\,{\delta\over\delta\alpha}]\,\,=\,\,\chi[-i{\delta\over\delta\rho},\,i\rho]\,.
\end{equation}
where $\rho$ is the charge density in the target wave function and $\alpha$ is defined by\cite{JIMWLK,cgc}
\begin{eqnarray}
&&\alpha^a(x,x^-)T^a\,\,=\,\,{1\over \partial^2}(x-y)\,
\left\{U^\dagger(y,x^-)\,\,\rho^{a}(y,x^-)\,T^a\,\,U(y,x^-)\right\},\nonumber\\
&&
U(x,x^-)\,\,=\,\,{\cal P}\,\exp\{i\int_{-\infty}^{x^-} dy^-T^a\alpha^a(x,y^-)\}\,.
\end{eqnarray}
I note that from the functional integral point of view this duality has been discussed earlier in \cite{dual}.

Consider the general expression for the $S$-matrix of a projectile with the wave function $|P\rangle$ scattering on a target with the wave function $|T\rangle$\cite{kl2}, where the total rapidity of the process is $Y$. The projectile is assumed to be moving to the left with rapidity $Y-Y_0$ (and thus has sizeable color charge density $\rho^-$), while the target is moving to the right
 with rapidity $Y_0$ (and has large $\rho^+$). We assume that the projectile and the target 
contain only partons with large $k^-$ and $k^+$ momenta respectively: $k^->\Lambda^-$ and $k^+>\Lambda^+$. 
The eikonal expression for the $S$-matrix reads
\begin{equation}
{\cal S}_{Y}\,\,=\,\,\int\, D\rho^{+a}(x,x^-)\,\, W^T_{Y_0}[\rho^+(x^-,x)]\,\,\Sigma^P_{Y-Y_0}[\alpha]\,,
\label{s}
\end{equation}
where $\Sigma^P$ is the $S$-matrix averaged over the projectile wave function
\begin{equation}
\Sigma^P[\alpha]\,\,=\,\,\langle P|\,{\cal P} e^{i\int dx^-\int d^2x\hat\rho^{-a}(x)\alpha^a(x,x^-)}\,|P\rangle\,.
\label{sigma}
\end{equation}
where $W^T[\alpha]$ is the weight function representing the target, 
which is related to the target wave function in the following way: for an arbitrary operator $\hat O[\hat\rho^+]$
\begin{equation}
\langle T|\,\hat O[\hat\rho^+(x)]\,|T\rangle\,\,=\,\,\int\, D\rho^{+a}\,\, W^T[\rho^+(x^-,x)]\,\,O[\rho^+(x,x^-)]\,.
\label{w}
\end{equation}
The field $\alpha(x)$ is the $A^+$ component of the vector potential in the light cone gauge $A^-=0$. This is the natural gauge from the point of view of partonic interpretation of the projectile wave function. 
In the formulae above we use hats to denote quantum operators. Note that the quantum operators 
$\hat\rho^{-a}(x)$ and $\hat\rho^{+a}(x)$
 do not depend on longitudinal coordinates, but only on transverse coordinates $x$. The "classical" variables $\alpha$ and $\rho^+$ on the other hand do depend on the longitudinal coordinate $x^-$. This dependence, as discussed in detail in \cite{kl} arises due to the need to take correctly into account the proper ordering of noncommuting quantum operators. Thus the ordering of the quantum operators $\hat\rho^+$ in the expansion of $\hat O$ in the lhs of eq.(\ref{w}) translates into the same ordering with respect to the longitudinal coordinate $x^-$ of $\rho^+(x^-)$ in the expansion of $O[\rho^+(x^-)]$ in the rhs of eq.(\ref{w}). 

As shown in \cite{kl} the functional $W^T[\alpha]$ cannot in general be interpreted as probability density, as it contains a complex factor. This factor - the Wess-Zumino term, ensures correct commutators between the quantum operators $\hat\rho^a$. In the present derivation we do not require an explicit form of this term, but the following property which is implicit in eq. (\ref{w}) is crucial to our discussion. The "correlators" of the charge density $\langle \rho^{a_1}(x_1,x^-_1)... \rho^{a_n}(x_n,x^-_n)\rangle$ do not depend on the values of the longitudinal coordinates $x^-_i$, but only on their ordering\cite{kl}.

Note that one can define an analog of $W^T$ for the wave function of the projectile via
\begin{equation}
\langle P|\,\hat O[\hat\rho^-(x)]\,|P\rangle\,\,=\,\,\int \,D\rho^{-a}\,\, W^P[\rho^-]\,\,O[\rho^-(x,x^-)]\,.
\label{wp}
\end{equation}
With this definition it is straightforward to see that $\Sigma^P$ and $W^P$ are related through a 
functional Fourier transform.
To represent $\Sigma$ as a functional integral with weight $W^P$ we have to order the factors of the charge density $\hat\rho^-$ in the expansion of  eq.(\ref{sigma}), and then endow the charge density $\hat\rho^-(x)$ with an additional coordinate $t$ to turn it into a classical variable. This task is made easy by the fact that the ordering of $\hat\rho$ in eq.(\ref{sigma}) follows automatically the ordering of the coordinate $x^-$ in the path ordered exponential. Since the correlators of $\rho(x,t_i)$ with the weight $W^P$ depend only on the ordering of the coordinates $t_i$ and not their values, we can simply set $t=x^-$. 
Once we have turned the quantum operators $\hat\rho$ into the classical variables $\rho(x^-)$, the path ordering plays no role anymore, and we thus have
\begin{equation}
\Sigma^P(\alpha)\,\,=\,\,\int\, D\rho^{a}\,\, W^P[\rho]\,\,e^{i\int dx^-\int d^2x\rho^{a}(x,x^-)\alpha^a(x,x^-)}.
\label{sw}
\end{equation}

We now turn to the discussion of the evolution.
The evolution to higher energy can be achieved by  boosting either the projectile or the target. The resulting $S$-matrix should be the same. This is required by the Lorentz invariance of the $S$-matrix.
Consider first boosting the projectile by a small rapidity $\delta Y$. 
This transformation leads to the change of the projectile $S$-matrix $\Sigma$ of the form
\begin{equation}
{\partial\over\partial Y}\,\Sigma^P\,\,=\,\,\chi^\dagger[\alpha,{\delta\over\delta\alpha}]\,\,\Sigma^P[\alpha]
\label{dsigma}
\end{equation} 
Substituting eq.(\ref{dsigma}) into eq.(\ref{s}) we have
\begin{eqnarray}
{\partial\over\partial Y}\,{\cal S}_{Y}&=&\int \,D\rho^{+a}(x,x^-)\,\, W^T_{Y_0}[\rho^+(x^-,x)]\,\,
\left\{\chi^\dagger[\alpha,{\delta\over\delta\alpha}]\,\,\Sigma^P_{Y-Y_0}[\alpha]\right\}\nonumber\\
&=&
\int \,D\rho^{+a}(x,x^-)\,\,\left \{\chi[\alpha,{\delta\over\delta\alpha}]\,\,
W^T_{Y_0}[\rho^+(x^-,x)]\right\} \,\,\Sigma^P_{Y-Y_0}[\alpha]\,.
\label{ds}
\end{eqnarray}
Where the second equality follows by integration by parts. 
We now impose the requirement that the $S$-matrix does not depend on $Y_0$
\cite{LL}. Since $\Sigma$ 
in eq.(\ref{s}) depends on the difference of rapidities, requiring that $\partial{\cal S}/\partial Y_0\,=\,0$ 
we find that $W$ should satisfy 
\begin{equation}
{\partial\over\partial Y}\,W^T\,\,=\,\,
\chi[\alpha,{\delta\over\delta\alpha}]\,\,W^T[\rho^+]
\label{dwt}
\end{equation}

Thus we have determined the evolution of the target eq.(\ref{dwt}) by boosting the projectile
and requiring Lorentz invariance of the $S$-matrix.
On the other hand the extra energy due to boost can be deposited in the target rather than in the projectile. 
How does $W^T$ change under boost of the target wave function? To answer this question
we consider the relation between $\Sigma$ and $W$  together with the evolution of $\Sigma$.
Referring to eqs.(\ref{sw}) and (\ref{dsigma}) it is obvious that multiplication of $\Sigma^P$ by $\alpha$ is 
equivalent to acting on $W^P$ by the operator $-i\delta/\delta\rho$, and acting on $\Sigma^P$ by 
$\delta/\delta\alpha$ is equivalent to multiplying $W^P$ by $i\rho$. Additionally, the action of 
$i\rho$ and $-i\delta/\delta\rho$ on $W^P$ must be in the reverse order to the action of 
$\delta/\delta\alpha$ and  $\alpha$ on $\Sigma^P$. This means that the evolution of the functional 
$W^P$ is given by
\begin{equation}
{\partial\over\partial Y}\,W^P\,\,=\,\,\chi[-i{\delta\over\delta\rho},i\rho]\,\,W^P[\rho]\,.
\label{dwp}
\end{equation}
Although eq.(\ref{dwp}) refers to the weight functional representing the projectile wave function, clearly the functional form of the evolution must be the same for $W^T$. Comparing eq.(\ref{dwt}) and eq.(\ref{dwp}) we find that the high energy evolution kernel must, as advertised,  satisfy the selfduality relation eq.(\ref{main}).
This is the main result.

The selfduality of the kernel is somewhat similar (although different in detail) to the duality symmetry of a harmonic oscillator Hamiltonian $p\rightarrow x,\ \ x\rightarrow -p$. One thus hopes that it may eventually be of help in solving the complete evolution equation, once it is derived.


\end{document}